\documentclass[a4paper, 10pt]{article}
\usepackage[utf8]{inputenc}
\usepackage[textwidth=470pt, textheight=650pt]{geometry}
\usepackage[natbib=true, maxcitenames=2, sortcites, sorting=ynt, style=apa, uniquename=false]{biblatex}

\makeatletter
\renewcommand{\maketitle}{\bgroup\setlength{\parindent}{0pt}
\begin{flushleft}
  \textbf{\LARGE{\@title}}
  \vspace{0.5cm}
  
  \@author
  \vspace{0.5cm}
  
  \@date
\end{flushleft}\egroup
}
\makeatother

\title{Women in academia: a warning on selection bias in gender studies from the astronomical perspective}
\author{M.~L.~L.~Dantas$^{1}$, E.~Cameron$^{2}$, Rafael~S.~de~Souza$^{3}$, A.~R.~da Silva$^{4}$, A.~L.~Chies-Santos$^{5}$, C.~Heneka$^{6}$, P.~R.~T.~Coelho$^{1}$, A.~Ederoclite$^{1}$, I.~S.~Beloto$^{1}$, V.~Branco$^{1}$, Morgan~S.~Camargo$^{1}$, V.~M.~Carvalho~de~Oliveira$^{1}$, C.~de~Sá-Freitas$^{7}$, G.~Gonçalves$^{1}$, T.~A.~Pacheco$^{1}$, Isabel Rebollido$^{8}$\\
\vspace{0.5cm}
$^{1}$ Instituto de Astronomia, Geofísica e Ciências Atmosféricas, Universidade de São Paulo, R. do Matão 1226, 05508-090, São Paulo, Brazil \\
$^{2}$ Curtin University, Kent Street, Bentley, Perth, 6102 WA, Australia \\
$^{3}$ Key Laboratory for Research in Galaxies and Cosmology, Shanghai Astronomical Observatory, Chinese Academy of Sciences, 80 Nandan Rd., Shanghai 200030, China\\
$^{4}$ Nicolaus Copernicus Astronomical Center, Polish Academy of Sciences, ul. Bartycka 18, 00-716, Warsaw, Poland \\
$^{5}$ Departamento de Astronomia, Instituto de Física, Universidade Federal do Rio Grande do Sul, Av. Bento Gonçalves, 9500, 91501970, Porto Alegre, R.S., Brazil \\
$^{6}$ Hamburg Observatory, University of Hamburg, Gojenbergsweg 112, 21029 Hamburg, Germany\\
$^{7}$ European Southern Observatory, Karl-Schwarzschild-Stra\ss e 2, 85748 Garching bei München, Germany \\
$^{8}$ Space Telescope Science Institute, 3700 San Martin Drive, Baltimore, MD 21218, USA 
}

\date{November 27, 2020}

\begin{document}

\maketitle

\begin{abstract}
    The recent paper by \citet{AlShebli2020} investigates the impact of mentorship in young scientists. Among their conclusions, they state that female protégés benefit more from male than female mentorship. We herein expose a critical flaw in their methodological design that is a common issue in Astronomy, namely ``selection biases''. An effect that if not treated properly may lead to unwarranted causality claims. In their analysis, selection biases seem to be present in the response rate of their survey (8.35\%), the choice of database, success criterion, and the overlook of the numerous drawbacks female researchers face in academia. We discuss these issues and their implications -- one of them being the potential increase in obstacles for women in academia. Finally, we reinforce the dangers of not considering selection bias effects in studies aimed at retrieving causal relations.
\end{abstract}

\vspace{0.5cm}

The many gender\footnote{For a broader view of gender identity see e.g. \citet[][]{Rubin2020}.} biases suffered by women in society are well established \citep[e.g.][]{Nguema2019}. Several studies have put in evidence the systematic challenges faced by women in societies around the world. Some of these manifest, but are not limited to, in remuneration for paid work: the gender pay gap  \citep{Madalozzo2010, Boll2019} and the impact of double shift work \citep{Vaananen2004, Gupta2008}; as well as in the burden of unpaid work\footnote{The unpaid work done by women is estimated to cost at least 10.8 trillion dollars yearly worldwide, according to the Oxford Committee for Famine Relief, OXFAM, 2020 technical report \citep[][]{Coffey2020}.} \citep[][]{Coffey2020}. Not to mention that non-white women or those who are from other minorities suffer from stronger versions of these biases, widening the gap between men and women \citep[e.g.][]{Armstrong2015, Phillip2018}; a problem that has been highlighted in particular with regard to barriers to entry in science, technology, engineering, and mathematics (STEM) careers \citep{McGee2016}. Furthermore, it seems that female scientists are more likely to have partners that are also male academics, whereas male scientists are mostly partnered with people that do not work outside of their home; consequently, even in a home with two academics, women in general have a higher load of household tasks \citep{Viglione2020}. Also, a funding gap seems to persist in science; women receive on average less grant funds, a phenomenon seen across all career levels \citep{Oliveira2019, Waisbren2008, Gordon2009, NAP2010}. The `Matthew effect' \citep{Merton1968}, meaning greater recognition is given to established scientists, magnifies such gaps. As lecturers, females suffer strong biases with students overwhelmingly favouring males \citep{MacNell2015}; females are expected to be more mothering than males and students tend to evaluate a female professor based on how well she is prepared in the classroom while male lecturers are expected to be charismatic and knowledgeable, which requires far less effort and time \citep[e.g.][]{Seierstad2012}. Therefore, for these reasons and others, women publish less in high-impact journals and, consequently, are less cited \citep[see e.g.][]{Lariviere2013, Dworkin2020}. Field socialisation processes often lead to the unjustified belief that these human differences do not interfere with academic success \citep[e.g.][]{Mattheis2020}.

It is in this context that the recent article by \citet*{AlShebli2020} has been published in Nature Communications. The authors propose to understand the impact of mentorship on young scientists via the statistical analysis of a large dataset of 3 million mentor-protégé pairs. As a measure of success, they make use of the number of publications and citations of mentors, a metric that knowingly inherits gender asymmetric bias favouring  male researchers \citep[e.g.][]{Lariviere2013}. Among their conclusions, they affirm that young female researchers benefit more from male mentors than female ones, concluding ``that current diversity policies promoting female-female mentorships, as well-intended as they may be, could hinder the careers of women who remain in academia in unexpected ways.'' Unfortunately, their analysis suffers from a fundamental methodologically flaw that astronomers, such as ourselves, will recognise as ``selection bias'' \citep[e.g.][]{norberg20022df, cameron2007galaxy, lauer2007selection, DeSouza2016, Dantas2021}. When working with a large catalogue of astronomical sources derived from a wide-field survey (e.g., stars, galaxies, black holes) one quickly discovers that many of the strongest apparent associations between the observable properties of these sources are the result of systematic biases in composition of the sample relative to the population as a whole; biases that without correction will obfuscate the astrophysical processes one aims at studying. 

The \citet{AlShebli2020} analysis is exposed to selection biases in at least two critical steps.  First, we note the very low response rate (8.35\%) to their survey on the nature of scientific collaborations, which serves as `validation' for their procedure to identify meaningful mentor-protégé pairs. With such a small response rate, the assumption of missingness entirely at random is unwarranted. In general, a smaller sample with higher response rate ($>80$\%) is recommended over a larger sample with low response rate \citep{evans1991good}. Journals should be cautious of accepting manuscripts presenting results based on very low response rates; and indeed some journals will have an explicit policy on the minimum acceptable response rate. For instance, a 2008 editorial in the American Journal of Pharmaceutical Education identifies a threshold of $>60$\% response rate for consideration \citep{fincham2008response}.

The second step in the \citet{AlShebli2020} study at which there is likely an exposure to selection bias  appears in the opportunistic sample of mentor-protégé pairs constructed from the Microsoft Academic Graph database. The sample includes only those individuals who remained as active researchers (at each given year since first publication). The post-doctoral years for researchers in many fields are characterised by short-term employment contracts, low salaries, and a high burden of travel and/or international relocation; as researchers from the field of organisational science have put it: ``postdocs seem to be trapped between their own ambitions and a lack of academic career opportunities'' \citep{van2016career}. It is thus no surprise that the challenges faced by women drive gender-biased differences in outcomes through this period \citep{mcconnell2018united,ysseldyk2019leak}. Although couched in the language of causal inference, the \citet{AlShebli2020} analysis fails to propose a graphical model (or equivalent) from which the decision on which variables to control for in the coarsened exact matching could be justified and from which the impact of likely selection effects could be understood \citep{bareinboim2012controlling}.

The timing of this publication with respect to the global SARS-CoV-2 pandemic is especially poignant considering, inter alia, the gender-biased impact of social isolation and remote working on research experiences. \citet{Myers2020a} showed that women suffered a stronger impact in the time dedicated to research when compared to male researchers during this period, particularly due to the impact of raising young children on their daily productivity. \citet{Staniscuaski2020} reinforce this analysis and discuss the imbalances of motherhood and fatherhood during the confinement period. The authors argued for changes in several policies (e.g. grant deadlines) as a way of mitigating the differential impact that the confinement had/has for men and women.

Many studies have shown the positive effects of a higher number of women across various environments, including in academia \citep[e.g.][]{Blau2010, Gaule2018}. \citealt{Stockemer2012} show that the higher is women's participation in the workforce, the higher their representation is in politics; and \citealt{Alexander2020} detail the positive impact on female citizens when being represented by female leaders. With this background we believe it is important that the concluding statements of \citeauthor{AlShebli2020} with regard to the impact of gender on mentor-protégé relationships be retracted.
Simply placing acknowledgements that a study may have some methodological limitations is not sufficient to protect against the misinformation effect of an analysis conducted with an intractably flawed design. The academic community has been making enormous efforts to mitigate the countless drawbacks women face -- e.g. female quotas in different academic positions, specific grants for women, which directly impact the establishment of additional female-female mentor-protégé relationships -- and their damping based on spurious evidence can cause serious effects in female representation in all areas. 

Selection biases are ubiquitous across data-driven studies, an effect that becomes even more significant in an era in which large datasets are readily available.  Astronomers have vast experience dealing with such effects for decades-long, as such, we provide a warning for better methodological designs, especially the ones aiming to derive causality claims.

\begin{refcontext}[sorting=nyt]
    \printbibliography
\end{refcontext}

\end{document}